# Final assembly, metrology, and testing of the WEAVE fibre positioner

Sarah Hughes[1a], Ellen Schallig[a,b], Ian J. Lewis[a], Gavin Dalton[a,c], David Terrett[c], Don Carlos Abrams[d], J. Alfonso L. Aguerri[e], Georgia Bishop[c], Piercarlo Bonifacio[f], Matthew Brock[a], Esperanza Carrasco[g], Kevin Middleton[c], Scott C. Trager[h], Antonella Vallenari[i]

[a]Dept. of Physics, Keble Road, University of Oxford, OX1 3RH, UK; [b]NOVA Optical Infrared Instrumentation Group at ASTRON, Oude Hoogeveensedijk 4, 7991 PD, Dwingeloo, The Netherlands; [c]RAL Space, Science and Technology Facilities Council, Rutherford Appleton Laboratory, Harwell Oxford, OX11 OQX, UK; [d]Isaac Newton Group, 38700 Santa Cruz de La Palma, Spain; [e]Instituto de Astrofisica de Canarias, 38200 La Laguna, Tenerife, Spain; [f]GEPI, Observatoire de Paris, Université PSL, CNRS, Place Jules Janssen, 92195 Meudon, France; [g]Instituto Nacional de Astrofisica, Optica y Electronica (INAOE), Mexico, [h]Kapteyn Instituut, Rijksuniversiteit Groningen, Postbus 800, 9700 AV Groningen, Netherlands, [i]Osservatorio Astronomico di Padova, INAF, Vicolo Osservatorio 5, 35122, Padova, Italy

**ABSTRACT**

WEAVE is the new wide-field spectroscopy facility[1,2,3] for the prime focus of the William Herschel Telescope at La Palma, Spain. Its fibre positioner is essential for the accurate placement of the spectrograph's 960 fibre multiplex. We provide an overview of the final assembly and metrology of the fibre positioner, and results of lab commissioning of its robot gantries. A completely new *z*-gantry for each positioner robot was acquired, with measurements showing a marked improvement in positioning repeatability. We also present the first results of the configuration software testing, and discuss the metrology procedures that must be repeated after the positioner's arrival at the observatory.

**Keywords:** fibre spectroscopy, WEAVE, WHT, robotic positioners, high multiplex spectroscopy, metrology, calibration

## 1. INTRODUCTION

WEAVE is the new spectroscopy facility[1,2,3] for the William Herschel Telescope (WHT) on La Palma. WEAVE has a multiplex of 960 optical fibres and a 2° field of view. It consists of four distinct subsystems: the prime focus system, the spectrograph system, the optical fibres, and the fibre positioner.

The prime focus system[4] consists of a six lens wide-field corrector with atmospheric dispersion compensator (ADC), an instrument rotator, and a new top end structure incorporating focus/tilt adjustment to the spider vanes. The ADC components comprise two air-spaced prismatic doublets which counter-rotate to correct for dispersion along the direction of the parallactic angle. The corrector prescription delivers <0.6" polychromatic images over a flat focal surface. The positioner is mounted to the field rotator, and the corrector is fixed to the top end structure.

WEAVE has three fibre observing modes: the MOS fibres, the mIFUs, and the LIFU. Each of the 20 mIFUs is a bundle of 37 fibres, whilst the LIFU has 547 central fibres and 8 bundles around the edge with 7 fibres for sky subtraction. The positioner has two 410 mm diameter field plates (plate A and plate B) located on either side of the central tumbler, that can rotate 180° between the two. Each field plate has an independent set of fibres, which allows the next field to be configured by the robots whilst the other field plate is being used for observations. This minimises the overhead time between observations.

The positioner uses pick-and-place technology to arrange the optical fibres in the focal plane according to the targets allocated in the field of view. Two robots move along *x, y, z,* and *θ*-stages to pick up individual fibres arranged around the edge of the field plate in retractors. A gripper system attached to the robots is used to place them in their desired position, within ±8 μm.

---

[1] sarah.hughes@physics.ox.ac.uk

The dual-arm spectrograph system[5] has continuous wavelength coverage (3660-9590 Å) in the low-resolution mode (R~ 5750) and coverage over several narrow regions (either 4040-4650 or 4730-5450, and 5950-6850 Å) in the high-resolution mode (R~ 21000). The spectrograph is not located on the top-end of the WHT, but on the Nasymth platform. This has significant advantages; such as improving the stability of the system when making observations, and reducing restrictions on the size of the spectrograph due to weight balance.

This paper will focus specifically on the fibres and the positioner, describing the final testing that was conducted prior to shipping.

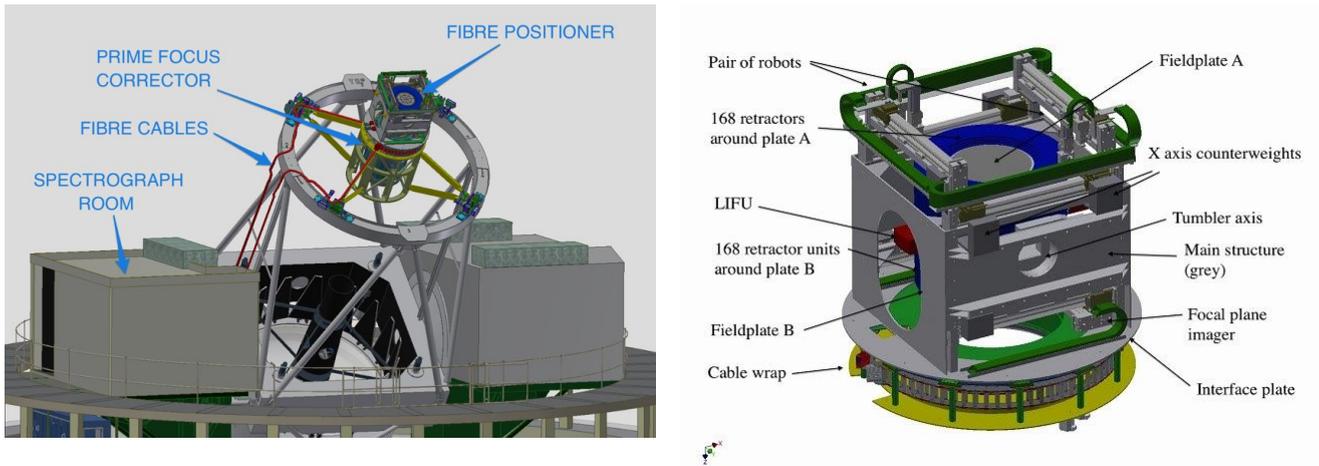

Figure 1: *On the left:* overview of the WEAVE instrument. The main components are the prime focus system, the optical fibres, the fibre positioner, and the spectrograph. *On the right:* a close-up of the fibre positioner, with the positioning robots at the top and the installed retractors in blue.

## 2. ROBOT GANTRIES AND AXES

Two sets of robot gantries are built into the positioner. The $x, y$-gantry on the bottom of the positioner, in the focal plane, holds the Focal Plane Imager (FPI) that surveys the field plate and the sky together to allow a determination of the mapping between the telescope focal plane and the field plates. On the top of the positioner, the two pairs of $y, z, \theta$-axes that share the same set of $x$-gantries are the two fibre positioner robots.

### 2.1 FPI-gantry

The gantry on the bottom of the positioner has only $x, y$ movements, and the two cameras of the FPI have fixed foci. The behaviour of this FPI-gantry must be mapped. The depth of focus of the sky-viewing camera, a Bigeye G-123B Cool thermo-electrically cooled camera with a pixel size of 6.45 μm, on the WHT is 18 μm[6]. This does not take into account any image degrading effects, such as seeing. As long as the vertical flexure of this gantry is smaller than this depth of focus, there is no information loss. The plate-viewing camera has a larger depth of focus due to extra lenses, and is therefore not the limiting factor.

We measured the shape of the plane patrolled by the FPI-gantry with a FARO Edge (a high-precision measurement arm). The Edge's tip rested in a measurement cone on the $y$-axis motor, and the motor was moved along the $y$-stage to specific positions. This was done for multiple positions along the $x$-gantry. Figure 2 shows the results after a coordinate transformation to internal positioner coordinates. The top left shows the data points taken in the $x, y$-plane, the top right shows the same points in the $x, z$-plane. The most interesting behaviour is in the $y, z$-plane, where measurements were made along the $y$-stage. The stage clearly flexes under the motor's weight. The deflection from the mean reading of 177.485 mm is ±0.02 mm. This is of the order of the depth of focus of the sky-viewing camera, and any seeing effects will be significantly larger than this.

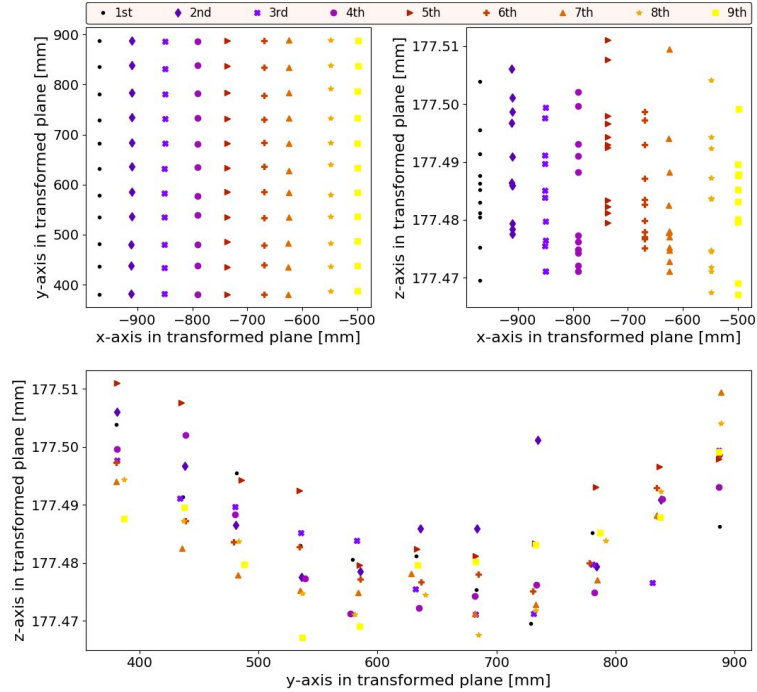

Figure 2: FPI-gantry measurements, around the centre of, and extending just past, the focal plane. Each colour is a measurement row along the *y*-stage, with a fixed position along the *x*-gantry. *Top left:* visualisation of the measurements in the *x, y*-plane. *Top right:* the measurements in the *x, z*-plane. *Bottom:* Measurements along the *y*-stage. The mean *z* of all the points is 177.485 mm, and the spread of most points is within ± 20 μm.

**2.2 Robot Positioner Gantries**

On the top of the positioner, the two positioner robots on *x, y, z, θ*-stages, each equipped with a gripper, have the task of positioning each fibre on the field plate. This must be within 8 μm of the desired position on the field plate, and with enough speed to stay within the observing time of the opposite plate. Each robot moves independently, and software ensures that they do not run into each other. To identify each robot, they have been given the names Morta and Nona[2]. Figure 3 shows an overview.

The repeatability of each positioner robot is a major contributor to the positioning budget of 8 μm. In the previous paper[7], we identified a significant problem with the *z*-stages. The repeatability was tested by measuring a stationary reference point: a backlit fibre on the field plate. After a move up and down along the stage, the camera takes a picture of the fibre, and centroiding software[8,9] finds its position in pixel coordinates. The results were strongly correlated in *x* and *y,* which are measured in image coordinates, not gantry coordinates. It appeared as if there was no temporal correlation. This last finding turned out to be an error, therefore we repeated the measurements with more data points. Figure 4 on the left shows a measurement run of 3669 data points, with the colour bar a proxy for the amount of time that has passed. There is a clear drift in *x* and *y* with time.

It was found that the specific design of these *z*-stages is responsible for the correlated movement in *x* and *y*. Rather than the motor moving along the stationary stage, as is the case for every other stage on the positioner, the opposite is true for the *z*-stages. Here the axis moves through the fixed motor, and the ends are not constrained. This, together with the cantilever of the gripper load along the *x*-axis, could account for the correlated behaviour.

---

[2] Nona and Morta are names from Roman mythology. They are two of the three sisters who represent destiny and control the thread of life.

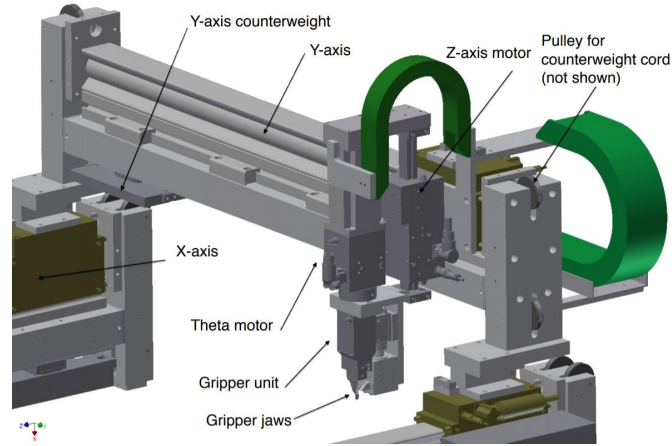

Figure 3: Each robot consists of its own $y$, $z$, $\theta$-axes, and shares the two $x$-axes with the other robot. The $\theta$-axis is aligned with the $z$-axis. The gripper consists of the gripper unit, custom-made jaws, and a camera.

The only way to fix this issue is to replace the $z$-motor stage with a different internal design. The new axes have significantly more bearings on each rail, a longer engagement of the motor on the rail, and adjustable preloads on both sides. The new design also allows a reduction in the distance from the gripper to the $z$-axis, therefore reducing the scatter in both $y$ and $x$. The new $z$-axis also includes a brake and magnetic preload. This was a new component developed by Schunk in collaboration with the project, and was not available at the original design stage.

Repeatability measurements with this new rail on Nona resulted in the plot on the right in figure 4. The spread in the 999 points is much lower, and will potentially be even smaller with more fine-tuning. Currently, the positioning repeatability of this $z$-gantry is 0.85 [px] × 4.4 [μm/px] = 3.7 μm (RMS). The overall repeatability of the robot axes is a total of 6.7 μm (RMS), with an $x$-$y$-repeatability of 5.5 μm (RMS)[6]. The drift from the right to the left is due to the gripper moving away from the $y$-axis as the $z$-motor heats up. This temperature effect will be calibrated out later, and will also have a positive effect on the repeatability of the $x$, $y$-stages. These measurements will have to be redone with Morta, to calibrate this effect out accurately on both robots.

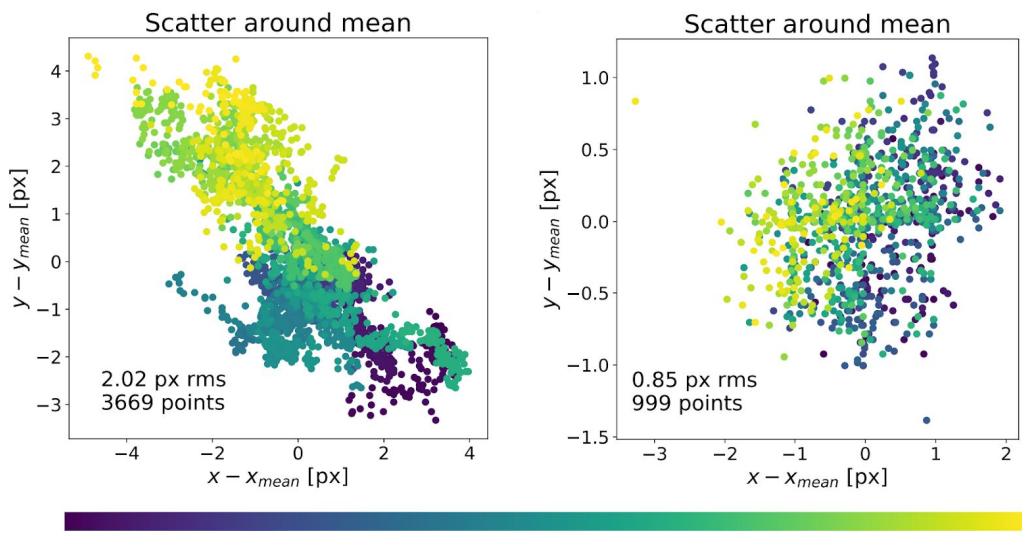

Figure 4: Effect of moving the gripper in $z$ only, away from and back to a stationary reference point. The spread of the $z$-measurements is in pixel units; each pixel is 4.4 μm wide[6]. The $x$- and $y$-values are in image coordinates, not gantry coordinates. The colours show a time series, with purple at the start, and yellow towards the end of the measurements. *On the left*: the original $z$-stage. *On the right*: the new $z$-stage (note the change of scale).

**2.3 FPI Camera**

The Basler plate-viewing camera is positioned to be parfocal with a plane 2.4 mm in front of the fieldplate to match the button/prism geometry. The BigEye camera is focused on targets in the night sky.

After aligning the FPI cameras, the unit was then installed onto its gantry to survey the grid of etched dots on the field plate and take images of each one. These surveys are vital for aligning the reference frame of the field plate to the coordinate frame of the sky. They are taken before a field is configured and again prior to observations beginning. However, when first completing these tests after the camera alignment, the images were oversaturated by the floodlights that illuminate the plate. This caused the software to fail at centroiding the grid point positions, and reducing the exposure time had little effect. To fix this, the aperture of the plate-viewing camera was reduced by installing a circular baffle, which prevented the images from saturating.

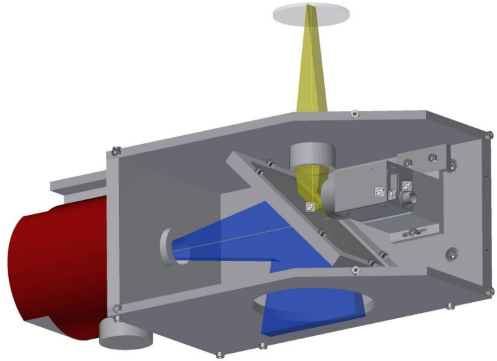

Figure 5: Diagram of the FPI, which contains a Basler plate-viewing camera (*in yellow*) focused on a plane 2.4 mm in front of the field plate, and a BigEye camera (*in blue*) that is focused on targets in the night sky. The FPI will be used for astrometry, mapping the reference frame of the field plate to the projection over the focal surface.

## 3. FIBRE INSTALLATION

**3.1 Retractor Assembly**

The installation of the optical fibres began in early December 2019. We successfully installed the 960 MOS fibers and 8 guide fibres (6 fibres per retractor, across 168 retractors per plate) required for plate A over the course of four months prior to the Covid-19 lockdown. The installations for plate B were completed on the 7th of August 2020 once lockdown restrictions eased. Figure 6 displays the complete installation of all retractors for plate A.

The installation of the optical fibres requires great care to complete. Whilst it is not a technically difficult process, the fibres only have a core diameter of 85 μm and 120 μm with cladding, which makes them extremely delicate. Any kinks or damages to the fibre will significantly impact the focal ratio degradation and throughput of the fibre. The prisms attached are only 1.5 mm square in size and are particularly susceptible to a shearing force, so physical contact with them must be avoided. As a result a single retractor takes about 1 hour to install, provided there are no issues with the movement of the fibres through the pulley system. The procedure to install a single retractor requires two people to complete.

Halfway through the installation of the retractors of plate A, a consistent issue of tier 1 fibres stalling when drawn out was discovered. Tier 1 refers to the lowest position of the fibres, with tier 3 being the highest. These tier 1 fibres performed as expected when the pulleys are tested before being bolted to the tumbler. This issue was produced when further retractors were installed in the neighbouring slots.

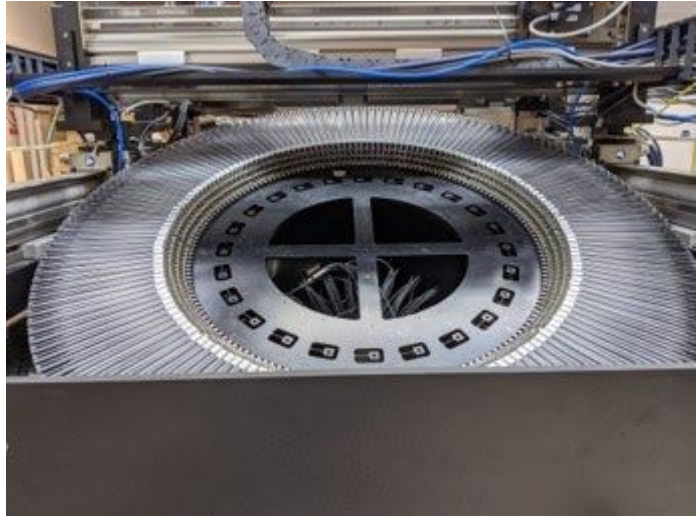

Figure 6: The complete set of 960 MOS fibres and 8 guide fibres for plate A, installed across 168 retractors which are bolted to the tumbler in a radial arrangement. Each retractor was assembled by hand, and this process took a period of 4 months to complete.

The retractors are arranged radially around the edge of the field plates, and neighbouring retractors are in contact with each other from the spare length box until the second pulley slot. The internal structure of the retractors is shown in figure 7. This contact, combined with the compact design spacing, led to pressure being applied on the tier 1 pulley system. The pressure prevented the pulley axle on one side from moving freely in the casing slots, stalling the movement of the pulleys and consequently the fibre itself.

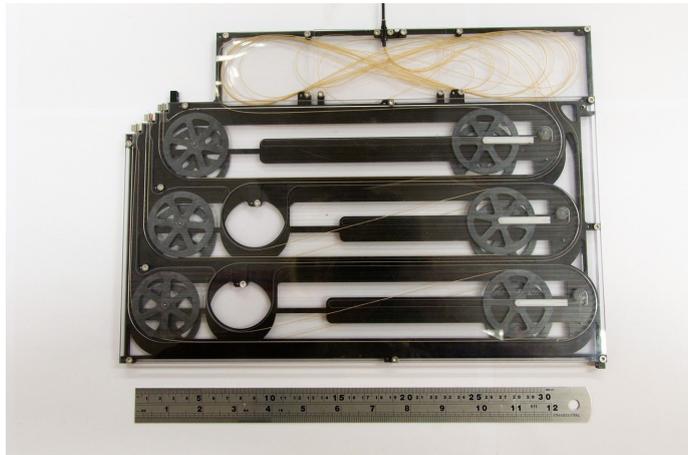

Figure 7: The internal structure of a single retractor, with the spare length box at the top. Each optical fibre is fed through two pulleys, which maintain a constant tension and prevent any twists forming. Each retractor stores 6 optical fibres, arranged in pairs across three tiers seen in the top left of the figure. Modifications had to be made to the outer casing during installations, due to applied pressure from neighbouring retractors. The casing was skimmed by 100 μm across the area covering the tier 1 pulleys, which resolved the issue.

To fix this issue, one side of the casing panels had to be removed for all retractors. A 100 μm thick layer was skimmed off the top of the area covering the tier 1 slot. This modification immediately rectified the issue by reducing the pressure that was being applied. Further testing of the fibre movement has not revealed any other problems at this stage.

## 3.2 Slit Assembly

To optimise the use of the detector spatially, all optical fibres need to be ordered linearly at the interface to the spectrograph. This is achieved by grouping the ends of the fibres into numbered metal casings called slitlets, which was completed prior to the fibres arriving at Oxford[10]. It is also important that the ordering of the fibres is recorded to match the spectra to the correct object.

There are 40 slitlets for both plate A and plate B, as well as 20 slitlets for the mIFUs. Each MOS slitlet contains 24 fibres, and each mIFU slitlet contains 37 fibres. The slitlets ensure that the polished ends of the fibres are aligned vertically and rest at a fixed distance in the interface to the spectrograph. The slitlets must then be secured into the slit itself. This is a curved, comb-like structure, with metal teeth for each slitlet to slot between. The slitlets are secured in place using screws that embed into the top of their casing.

To prevent any reflection or refraction due to air at the interface between the slitlets and the glass window of the slit, a layer of index-matching gel is required. The gel must be applied evenly to the front of each slitlet before it is inserted into its correct slot. It is crucial that the slitlet is firmly pressed against the glass window as it is being secured in place, otherwise air bubbles will be introduced into the system.

Each slitlet is inspected closely with a camera to identify any air bubbles that cover the surface of the fibres. If any are present, the slitlet will be removed and cleaned and a new layer of gel will be applied before it is secured in place again. Figure 8 shows a MOS slit with about 75% of the slitlets successfully installed. The entire process took 2 weeks to complete. The LIFU slit had already been assembled by NOVA.

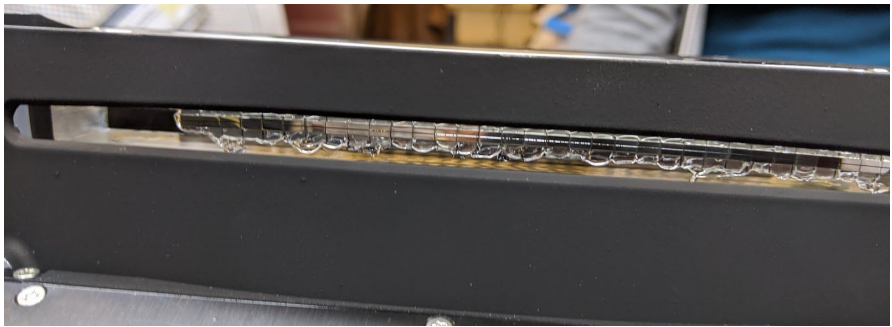

Figure 8: The MOS plate B slit, with about 75% of all slitlets successfully installed. Index-matching gel has been applied to the surface of the slitlets to prevent any reflection at the boundary between the slitlets and the slit window. The fibres are back-illuminated to identify any air bubbles in the gel across their surface.

## 4. TUMBLER METROLOGY AND PLATE CALIBRATIONS

### 4.1 Tumbler Metrology

A crucial requirement for WEAVE is that the field plates are co-focal. This means that when each plate is being used for observations, both plates lie in the same position within the focal plane to 15 μm. This requirement corresponds to the depth of focus at f/2.7. If this error budget is exceeded then some fibres will always be out of focus during observations.

The metrology of the tumbler was initially completed in 2018, prior to any retractor installations. The complete set of retractors placed approximately 200 kg of additional weight onto the tumbler, therefore the metrology must be re-measured.

Each plate was tumbled into the focal plane and a sparse grid of points across the surface was measured using a dial gauge mounted on a calibration bracket attached to the FPI gantry. A diagram demonstrating this arrangement is shown in figure 9. The dial gauge was moved with the gantry by hand to a set of locations across the field plate.

The measurements determine any tilt along the *x*- and *y*-directions, as well as an estimate of the position of the central tumbler beam. The results showed that the field plates were tilted in opposite directions forming a cone-like shape, with a tilt of 500 µm in plate A and 200 µm in plate B.

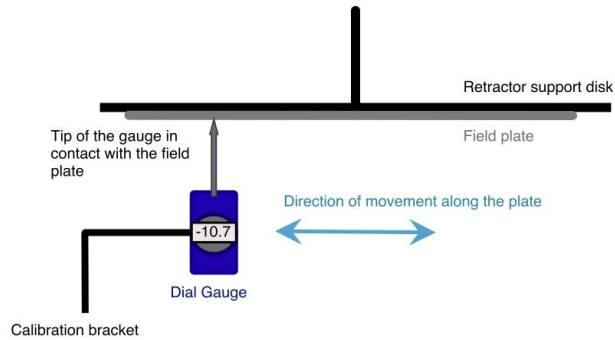

Figure 9: Arrangement of the dial gauge with respect to the field plates and the tumbler. The dial gauge is attached to the calibration bracket, which can be moved by hand to specific *x, y*-positions along the gantries. Height measurements can then be made along the field plates to identify any tilt that must be corrected for.

The central tumbler beam had shifted by 200 µm from its previous position and had to be adjusted. Due to the tilt shown across both plates, the axis needed to be lowered by 80 µm. This was difficult to correct for, and was achieved by very slowly loosening and re-tightening the bolts around the tumbler beam, using gravity to move it downwards in a controlled procedure. A diagram of the measured structure is shown in figure 10.

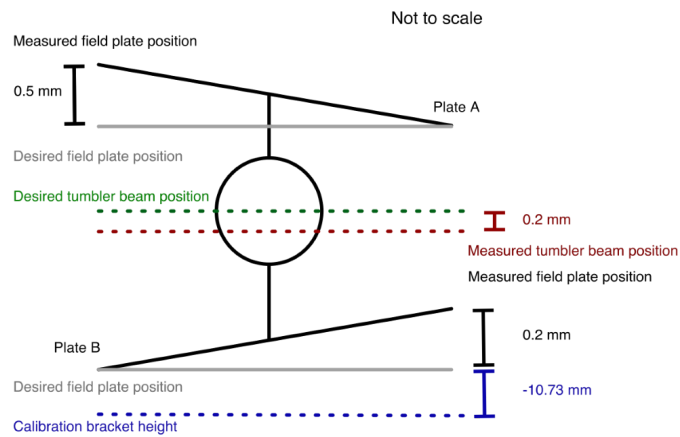

Figure 10: A diagram demonstrating the arrangement of the tumbler after the fibre installations were completed. There is an obvious cone structure formed by the two field plates which can only be corrected for using positive shims. The central tumbler beam has also shifted by 200 µm from the previous measurement and must be shifted to ensure the plates are co-focal in the observing position.

To correct for the tilts in both field plates, shims were inserted beneath each of them, with a total thickness of 700 µm distributed across them according to their tilt. This significantly reduced the tilt when in the observing position so that it is approximately flat. However, from the measurements a 100 µm tilt along the field plate when in the opposite orientation is expected.

The position of the tumbler beam will need to be re-measured with the same procedure once the positioner arrives at the observatory. It is likely to have changed during transportation, as well as due to the difference in temperature compared to when the original measurements were taken.

## 4.2 Field Plate Calibrations

Since both field plates are now arranged to be co-focal in the observing position, they are not guaranteed to be co-planar in the frame of the positioning robots. Tilt and displacement of the field plates relative to the robots contribute to the fibre position error budget and therefore need to be mapped. Measurements of the plate's $z$-height were taken in robot coordinates, using each robot at set coordinate positions which span the whole field. To protect the gripper jaws of each robot, the robots were not driven onto the plate directly. Instead they were moved to a set nominal height above the plate, and gauge blocks were used to measure the difference between this nominal height and the field plate.

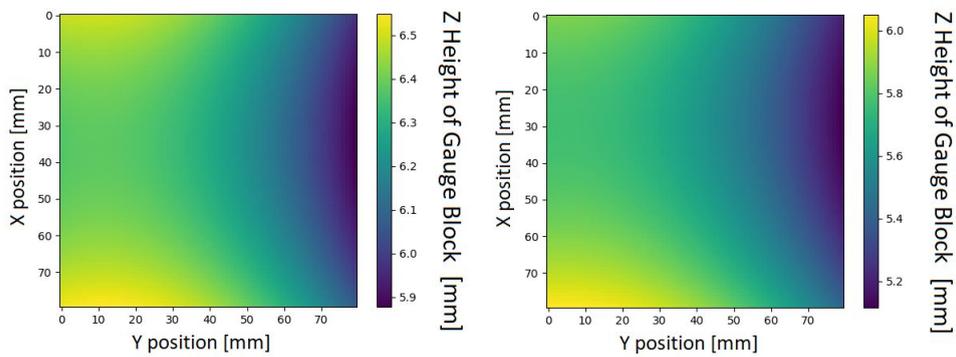

Figure 11: A contour map of the $z$-height measured for field plate A using robots Nona (*left*) and Morta (*right*). Both measurements were made using gauge blocks at set coordinates across the plate.

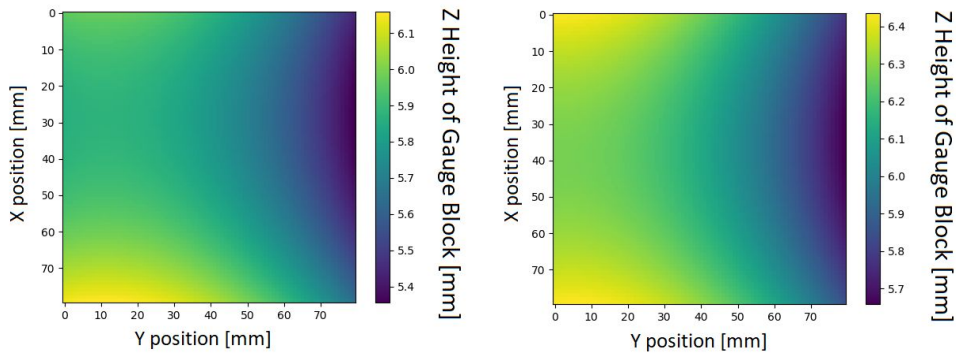

Figure 12: A contour map of the $z$-height measurements for field plate B using the robots Nona (*left*) and Morta (*right*). Both measurements were made using gauge blocks at set coordinates across the plate.

It is impossible to entirely isolate the warping of the field plate from effects due to the robot gantries using this method of measurement. At best, you can either infer the difference between both plates, or the difference between the two robots. These measurements need to be repeated on-site once the positioner arrives at the observatory, as the difference in temperature is also likely to change the metrology of the field plates and the robot gantries.

Figures 11 and 12 display the final $z$-maps for each robot and each field plate. There is a clear tilt in the $y$-direction of 140 μm across both plates, which will be corrected for during fibre placement by incorporation of a surface map into the positioning software.

## 4.3 Retractor Tier Heights

As with the field plates, the variation in $z$-height between the retractor porches must be measured to reduce errors during fibre placement, not least because the field plate is now tilted by the shims with respect to the retractor support plate. It was not possible to use gauge blocks due to the high risk of damaging neighbouring fibres. The measurement values were taken by turning off the $z$-axis brakes and lowering the robot by hand until the gripper jaws touched the porch. The data was read directly from the value of the encoders. To prevent a collision between the robot and the fibres, the tier 3 fibres were first placed onto the plate at a position 14° from their pivot position in park. This angle was chosen to ensure that the fibres are clear of the porches. The movement procedure required is described further in Section 5.

This method is only possible for the highest tier porches, due to the risk of hitting fibres on other tiers when attempting to measure tiers 1 and 2 directly. The retractor porches were designed to be 10 mm apart vertically, so the measurements made using tier 3 can be extrapolated downwards. Any tilt of individual porches on tier 3 can be easily identified by comparing the difference in height across two measurements. The extrapolation does not identify this possible tilt across the lower tier porches, which must be checked by visual inspection and may also be found when parking a fibre. In most cases the tilt across a porch can be easily fixed by adjusting the screw that secures it in place.

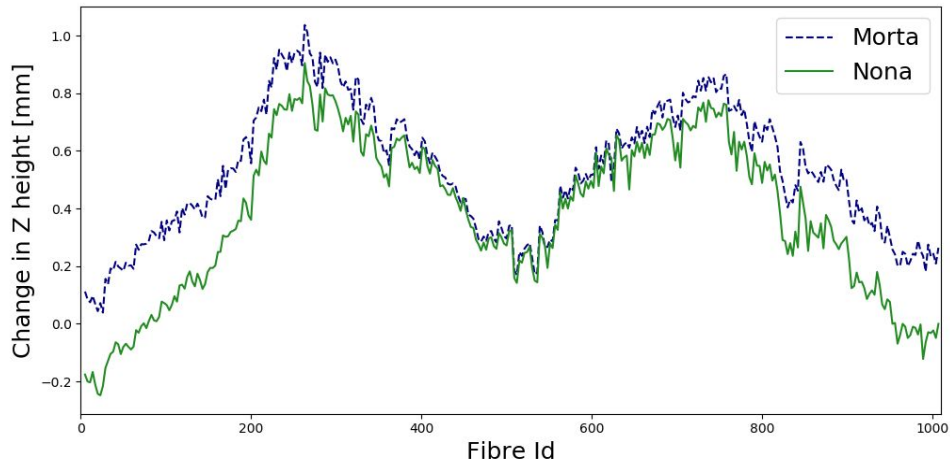

Figure 13: The change in nominal $z$-height for each fibre position on the tier 3 retractor porches for plate A. The periodic change in $z$-height is due to shim-induced tilt in the field plate, and the tilt differences between the $x$-stages as the robots move around the edge of the field. The offset between Nona and Morta at the furthest points across the field are due to the rest height of each robot's $y$-stage on top of each $x$-axis.

The change in $z$-height across all retractors for plate A is shown in figure 13. There is a clear periodic change around the edge of the field, which corresponds to shim-induced tilt in the field plate and tilt differences between the two $x$-stages, as the robots gradually move around the edge of the field. The periodic pattern is consistent for both Nona and Morta, which is expected as they share the same $x$-gantry. The offset between them is a result of differences between how the two $y$-stages rest on the $x$-gantry. From the offset, it is likely that the $y$-gantry of Nona rests slightly higher on one side of the $x$-gantry. Due to time constraints, the offset between Nona and Morta is taken to be fixed for each retractor position, meaning that the plate B tier heights only needed to be measured with one robot.

# 5. FIBRE MOVEMENT TESTS

## 5.1 Initial Movements

Each optical fibre has a metal button at the end, which has a central vane where the robot gripper jaws can pick them up from. Attached to this button is a prism, which directs light down the length of the fibre, and a magnet, which secures the button to the field plate. This setup is shown in figure 14.

A software script was developed independently from the program that is used to run full field configurations, for the purpose of engineering tests. The development of the script was used exclusively for all first movements of the fibres onto the field plate. Due to assembly of the retractors by hand, many of the fibres were not in the optimal position on their respective porches. The nominal park position is when the button vane is radial to the centre of the field and in line with its pivot point. It was decided that the procedure to correct the positions would be combined with measuring the $z$-heights of the retractor porches.

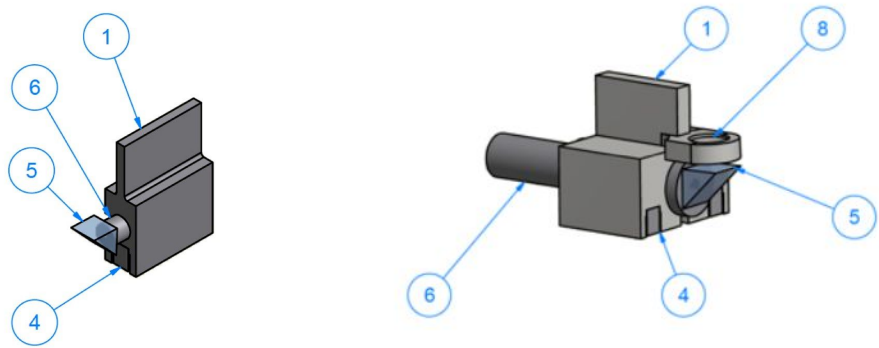

Figure 14: Schematic of the MOS (*left*) and mIFU (*right*) magnetic fibre buttons made of stainless steel. In each drawing, part 1 is the button vane, 4 is a permanent magnet, 5 is a prism, and 6 is the fibre ferrule. Part 8 is the separate field lens for the mIFU button.

To reset the park positions, Nona was moved to the nominal park position with the gripper jaws open. The robot was then slowly lowered to the nominal pick-up height. For any button out of position, Nona's coordinates were adjusted in small increments to prevent the gripper jaws from hitting the button vane.

At the nominal pick-up height, the button's orientation was determined using the centroiding software[8]. Based on its position, the robot's coordinates were changed to align the gripper jaws to the centre of the button vane, at the correct radial distance. The fibre was then placed radially on the plate, before being moved to a position ±14° from its pivot position, depending on which side of the porch the fibre is parked. This ensures that the fibre is clear of its respective porch.

All initial movements of the fibres were made with Nona only, as further calibration of Morta's $\theta$-axis and gripper camera were still required at that stage. However, the $z$-height measurements were still taken using both robots to determine the offset between them. The robots were first moved over the park position listed in the database for a particular fibre. With the gripper jaws open, the robot was slowly moved to the nominal pick-up height. In many cases, minor adjustments had to be made to the robot's orientation to prevent the gripper jaws from hitting the vane of the button.

After the tier $z$-height measurements are taken, the fibre is placed back in its park position and re-imaged to record any shift in its orientation. Following this, no manual adjustments are required to move the fibre again, as each button is in a known position with a known orientation.

## 5.2 Tier 1 Fibre Movement

In the initial setup of the robots, the axes movement was programmed to have the $x, y$ and $\theta$-axes move simultaneously. The $z$-axis was originally kept as a distinct axis that could move independently to movement of the other axes. However, the height of the tier 1 porch is approximately in line with the height of the field plate, which means that the pull-out move must take place in $x, y$ and $z$, simultaneously. The low-level software has therefore been reconfigured to incorporate each $z$ in the respective kinematic robot with fully synchronised 5-axis moves.

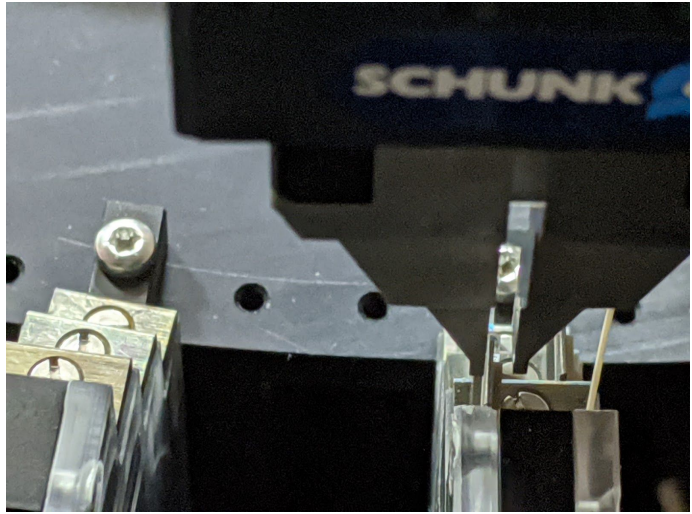

Figure 15: A visual inspection of the gripper jaws relative to the vane of the fibre button. The jaws should be centred on the button to move down to the nominal pick-up height and image the fibre. This is the first step in the procedure for picking up a fibre that has not been moved yet.

A condition has been applied to cause the tier 1 fibres to be drawn out in a two-stage movement. It is moved at a reduced speed to an intermediate position above the plate perimeter before being moved into the field. The reverse is true when parking them, and this procedure has been tested on all tier 1 fibres.

## 5.3 Prism Imaging

A key feature of the positioner software[11] is a check against the known locations of all other fibres in the field, when placing a fibre to prevent any collisions. Each fibre has a software buffer of about 50 µm around its boundary that prevents other fibres being placed too close. However, this buffer is currently set using a default outline of the button and the prism. This does not account for any variations in the relative alignment of the prism to the front edge of the button.

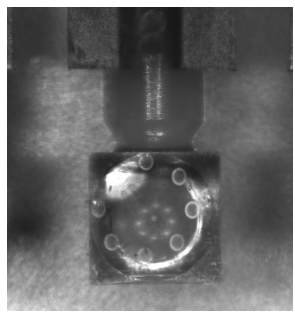

Figure 16: Image of the prism for a single optical fibre, taken with Nona's gripper camera. This image will be used to define the boundary of the prism within the configuration software, and will also identify any prisms that have been damaged. The structure at the top is the front of the button, the square piece is the prism, and the section that connects them is the fibre ferrule. The rings in the image around the centroid are reflections of the floodlights, which are LEDs arranged around the edge of the camera aperture.

To correct this, the boundary of each prism must be defined using direct images. It will also identify any prisms that have been damaged during the installation and testing process. An example of these images is shown in figure 16.

The imaging process was completed in conjunction with tests of the final configuration software, by generating a field with all fibres in a selected tier. The field placed each fibre in a ring arrangement. This is shown in figure 17. These images can be taken independently from this configuration, and will need to be done following the repair of any fibre. The images were taken with the gripper camera of Nona with the spotlight on, for consistent illumination across the field. This was repeated for all three tiers across both plates. This imaging process will need to be repeated when any repairs are made to the optical fibres.

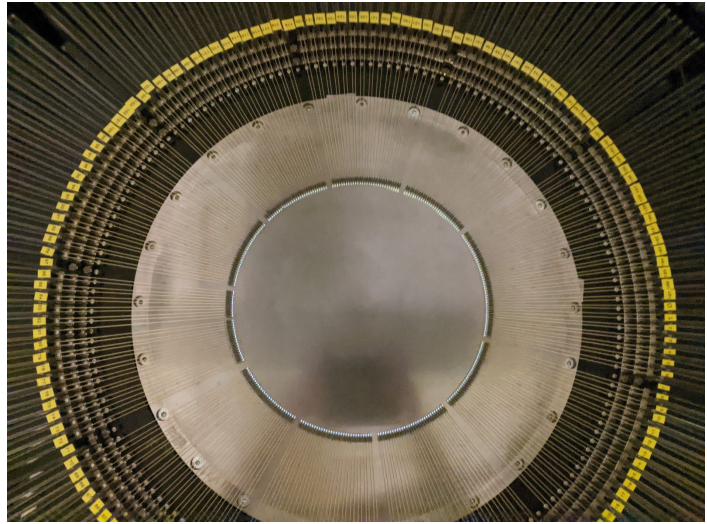

Figure 17: The ring configuration used to image the prism of each fibre on plate B and the first test of the final configuration software that will be used during observations. Every fibre in tier 3 has successfully been placed radially on the plate, with gaps corresponding to the location of the mIFUs.

### 5.4 Fibre Crossings

To complete configurations that meet the target separation requirements, the limitations of optical fibres crossing within the field must be understood. All previous movement tests have only placed the fibres radially. This step requires careful documentation of the order that the fibres were placed on the plate. If either robot attempts to move a fibre that is underneath another in the field, then a large amount of damage may be caused to fibres placed nearby.

The fully automated configuration software has been tested previously with the ring configuration. However, for this test each fibre movement was conducted using single commands so that the crossings could be checked at each stage. This test field is shown in figure 18. The order of the fibre placement was tightly controlled and is easily reversible. The order was generated using the configuration software, allowing the process to be repeated as a field. Notably, this has only been completed using one octant of the optical fibres on plate A. Further testing is required to reliably generate a complete field that contains crossings, and places them within the allowed time frame.

This test did not highlight any potential limitations of the fibre crossings at this stage. The tier 1 fibres are placed from park before the upper tiers, as these would cross at much higher $z$ close to the edge of the field. Further restrictions on the placement order may be needed, which may impact the field configuration order.

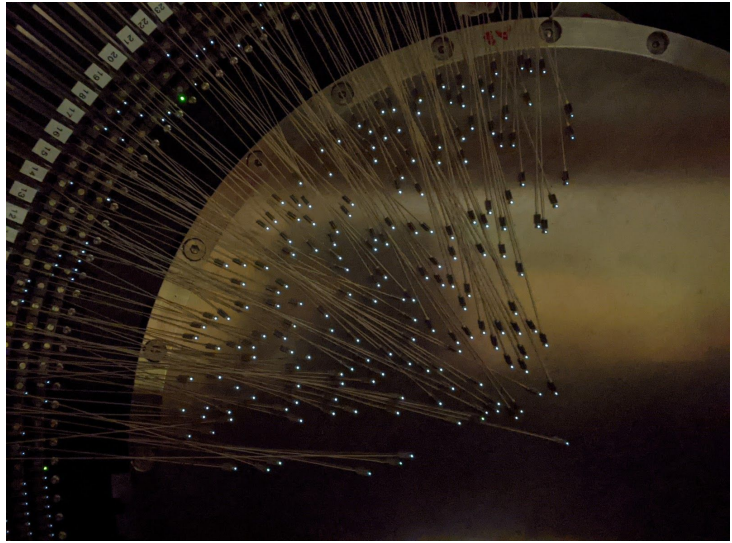

Figure 18: The first configuration to test the crossing of optical fibres, initially completed using a list of individual movements, before being repeated by the final configuration software. The field contains one octant of the plate A fibres across all three tiers.

## 6. PACKING AND SHIPPING

On the 12th November 2020, the positioner was carefully moved out of the lab in Oxford and placed in a container ready for transport to La Palma, shown in figure 19. The process of moving the positioner out of the lab was a difficult task, as the positioner and its base weighs approximately 1.5 tonnes. Attached to the positioner is a group of four wooden trolleys that contains the length of the optical fibres, coiled in figures of eight. A large team was required to gradually lift the positioner and construct a railway track to move it out of the building.

At the time of writing this paper, the positioner is still in transit, and is expected to arrive on-site by mid-December 2020. As a result, the commissioning period of WEAVE will begin in January 2021.

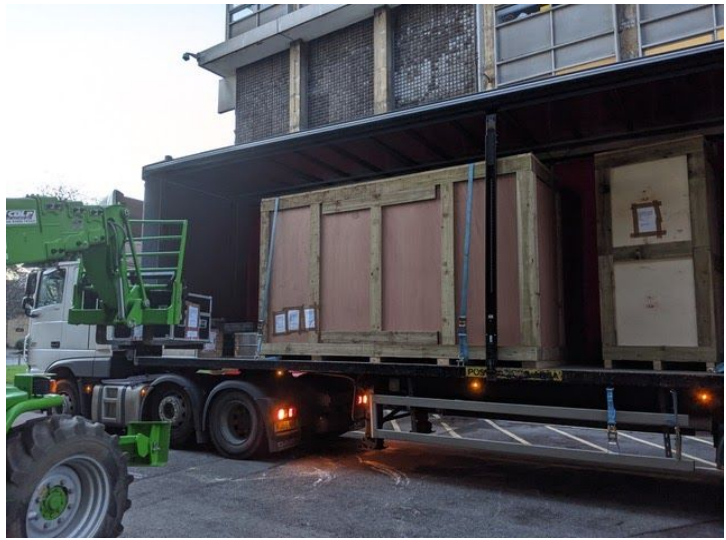

Figure 19: Image of the completed positioner, stored in a wooden crate before departing from Oxford. The crates were placed in a shipping container and are in transit to La Palma at the time of writing this paper.

# 7. FUTURE OUTLOOK

The construction of the WEAVE fibre positioner is now finished, displayed in figure 20, with a complete population of optical fibres. It was extensively calibrated and tested in the lab, and is in the process of being shipped to the observatory. After its arrival, several of the calibration measurements, such as the tumbler metrology and the field plate $z$-heights, will need to be repeated to check the alignment. The positioner will then be placed at prime focus, where on-sky commissioning can begin. Further testing of the configuration software is needed, to ensure that each field can be completed in the allotted observing time.

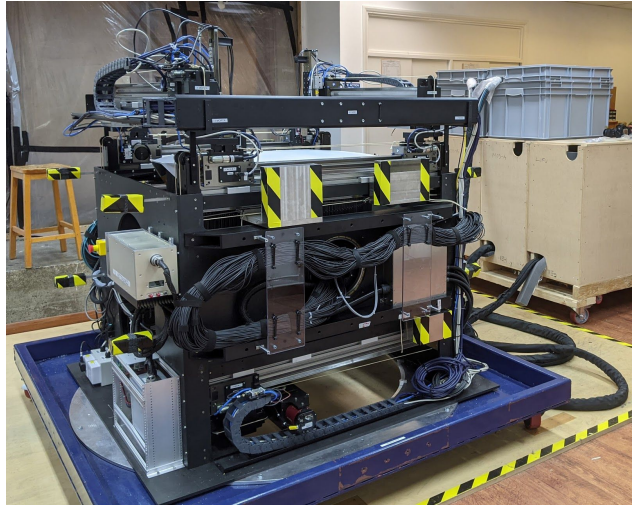

Figure 20: Final image of the WAVE fibre positioner, prior to the assembly of the covers. The wooden containers on the right hold the remaining length of the optical fibres that will be attached to the telescope, leading to the spectrograph.